\documentclass[aps,prd,twocolumn,nofootinbib,showpacs]{revtex4}

\usepackage{graphicx,color}
\usepackage[colorlinks=true, linkcolor=blue, citecolor=blue, urlcolor=blue]{hyperref}
\usepackage{color}

\begin{document}

\title{Scale-Fixed Predictions for $\gamma + \eta_c$ production in  electron-positron collisions at  NNLO in perturbative QCD}

\author{Huai-Min Yu$^{1}$}

\author{Wen-Long Sang$^{2}$}

\author{Xu-Dong Huang$^1$}

\author{Jun Zeng$^1$}

\author{Xing-Gang Wu$^1$}
\email[email:]{wuxg@cqu.edu.cn}

\author{Stanley J. Brodsky$^3$}
\email[email:]{sjbth@slac.stanford.edu}

\address{$^1$ Department of Physics, Chongqing University, Chongqing 401331, P.R. China}
\address{$^2$ School of Physical Science and Technology, Southwest University, Chongqing 400700, P.R. China}
\address{$^3$ SLAC National Accelerator Laboratory, Stanford University, Stanford, California 94039, USA}

\date{\today}

\begin{abstract}

In the paper, we present QCD predictions for $\gamma+\eta_{c}$ production at an electron-positron collider up to next-to-next-to-leading order (NNLO) accuracy without renormalization scale ambiguities. The NNLO total cross-section for $e^{+}+e^{-}\to\gamma+\eta_{c}$ using the conventional scale-setting approach has large renormalization scale ambiguities, usually estimated by choosing the renormalization scale to be the $e^+ e^-$ center-of-mass collision energy $\sqrt{s}$.  The Principle of Maximum Conformality (PMC) provides a systematic way to eliminate such renormalization scale ambiguities by summing  the nonconformal $\beta$ contributions into the QCD coupling  $\alpha_s(Q^2)$.  The renormalization group equation then sets the value of  $\alpha_s$ for the process. The PMC renormalization scale reflects the virtuality of the underlying process, and the resulting predictions satisfy all of the requirements of renormalization group invariance,  including renormalization scheme invariance. After applying the PMC, we obtain a renormalization scale-and-scheme independent prediction, $\sigma|_{\rm NNLO, PMC}\simeq 41.18$ fb for $\sqrt{s}$=10.6 GeV. The resulting pQCD series matches the series for conformal theory and thus has no divergent renormalon contributions. The large $K$ factor which contributes to this process reinforces the importance of uncalculated NNNLO and higher-order terms. Using the PMC scale-and-scheme independent conformal series and the $\rm Pad\acute{e}$ approximation approach, we predict $\sigma|_{\rm NNNLO, PMC+Pade} \simeq 21.36$ fb, which is consistent with the recent BELLE measurement $\sigma^{\rm obs}$=$16.58^{+10.51}_{-9.93}$ fb at $\sqrt{s} \simeq 10.6$ GeV. This procedure also provides a first estimate of the NNNLO contribution.

\pacs{12.38.Bx, 13.66.Bc, 14.40.Lb}

\end{abstract}

\maketitle

\section{Introduction}

Processes involving the production of heavy quarkonium are important for testing Quantum Chromodynamics (QCD) as well as the effective theory of Nonrelativistic QCD (NRQCD)~\cite{Bodwin:1994jh}. The framework of NRQCD factorization theory allows the non-perturbative dynamics involving the binding of the heavy quark-antiquark pair in quarkonium to be factored into universal NRQCD matrix elements which can be extracted and fixed via a global fitting of experiments involving heavy quarkonium production. The remaining `hard' contribution involving higher momentum transfers is then perturbatively calculable. Thus reliable calculations of quarkonium production and its decay now appear viable.

The NRQCD approach has been successfully applied to a number of quarkonium processes, but many challenges and puzzles have remained. At present, most NRQCD results have been done at the next-to-leading order (NLO) level. The next-to-next-to-leading order (NNLO) and higher-order calculations are much more difficult. Thus it is important to test a variety of NNLO predictions before drawing any definite conclusion on the general applicability of NRQCD, especially since  the $K$ factors which appear in heavy quarkonium production processes can be very large.

The $\eta_c$ production via the process $e^{+}+e^{-}\to\gamma+\eta_c$ is an important charmonium production process which can be precisely measured at high-energy, high-luminosity electron-positron colliders. For example, the Belle II experiment is expected to produce a sizable number of $\gamma+\eta_c$ events in the near future, which can be used to make precise comparisons with theoretical predictions. This heavy quarkonium production process has been calculated in pQCD up to NNLO level~\cite{Shifman:1980dk, Sang:2009jc, Li:2009ki, Chen:2017pyi, Chung:2019ota}. However, the NNLO calculation performed by Chen, Liang and Qiao~\cite{Chen:2017pyi} displays both a large $K$ factor and large renormalization scale uncertainties. As a cross check, and also as important step forward, we shall recalculate the cross section of $e^{+}+e^{-}\to\gamma+\eta_c$ at the NNLO level,  make a detailed discussion on how to eliminate the unnecessary renormalization scale ambiguities, and present a first prediction of the NNNLO contribution.

A physical observable, corresponding to an infinite-order pQCD approximation, doesn't depend on the choice of renormalization scale. If the perturbative coefficients and the strong coupling constant $\alpha_s$ are not well matched at a fixed order, as is the case of conventional scale-setting approach in which the renormalization scale is simply guessed, one finds significant renormalization scale-and-scheme ambiguities; cf. the reviews~\cite{Wu:2013ei, Wu:2014iba, Wu:2019mky}. Any dependence of pQCD prediction on the choice of renormalization scheme violates a fundamental principle of the renormalization group. In fact, predictions based on  conventional scale-setting approach are even incorrect for Abelian theory -- Quantum Electrodynamics (QED); the renormalization scale of the QED coupling constant $\alpha$ can be set unambiguously by using the Gell-Mann-Low method~\cite{GellMann:1954fq}. It is thus essential to use a rigorous scale-setting approach in order to achieve reliable and precise scale-and-scheme independent fixed-order pQCD predictions.

The Principle of Maximum Conformality (PMC)~\cite{Brodsky:2011ta, Brodsky:2012rj, Brodsky:2011ig, Mojaza:2012mf, Brodsky:2013vpa} provides a rigorous approach to renormalization scale setting, extending the BLM method~\cite{Brodsky:1982gc} to all orders in pQCD. The purpose of the PMC is not to find an `optimal' renormalization scale, but to achieve renormalization scale independent pQCD prediction. This can be achieved by systematically and rigorously determining the effective value of $\alpha_s$ of the process based on the renormalization group equation (RGE), which is free of renormalization scale dependence. Following this way, the scheme-dependent non-conformal $\{\beta_i\}$-terms~\footnote{There are $\{\beta_i\}$-terms for the renormalization of the heavy quark' mass and wave function, which are irrelevant to the renormalization of $\alpha_s$ and should be kept as conformal coefficients. Those terms can be adopted for fixing a more accurate quark mass itself, which is however out of scope of the present paper.} are eliminated in the pQCD series, which matches the corresponding conformal series. Thus after applying the RGE, the resulting pQCD prediction is also independent of the choice of the renormalization scheme. The commensurate scale relations~\cite{Brodsky:1994eh} which relate PMC predictions for different observables among each other also ensure the scheme independence. Moreover, the PMC procedure reduces in the $N_C\to 0$ Abelian limit to the Gell-Mann-Low method~\cite{Brodsky:1997jk}. Thus the PMC eliminates renormalization scale-and-scheme ambiguities simultaneously, satisfying the principles of renormalization group invariance~\cite{Brodsky:2012ms, Wu:2018cmb}. In addition, since the $\{\beta_i\}$-terms have been removed, the divergent renormalon terms like $\alpha_s^n \beta_0^n n!$ disappear and a convergent perturbative series can be naturally achieved.

In the paper, we shall adopt the PMC single-scale approach~\cite{Shen:2017pdu} for our analysis. The PMC scale can be interpreted as the effective momentum flow within the production process; its value displays stability and convergence with increasing order in pQCD, and any residual scale dependence due to unknown higher-order terms is highly suppressed.

The remaining parts of the paper are organized as follows: In Section II, we give the calculation technology for the total cross section of $e^{+}+ e^{-}\to \gamma+\eta_c$ up to NNLO level. We present the numerical results in Section III, and Section IV is reserved for the summary.

\section{Calculation Technology}

In the following, we shall give a brief description on the calculation techniques, and then present the numerical expressions up to the NNLO level.

\begin{figure}[htb]
\centering
\includegraphics[width=0.48\textwidth]{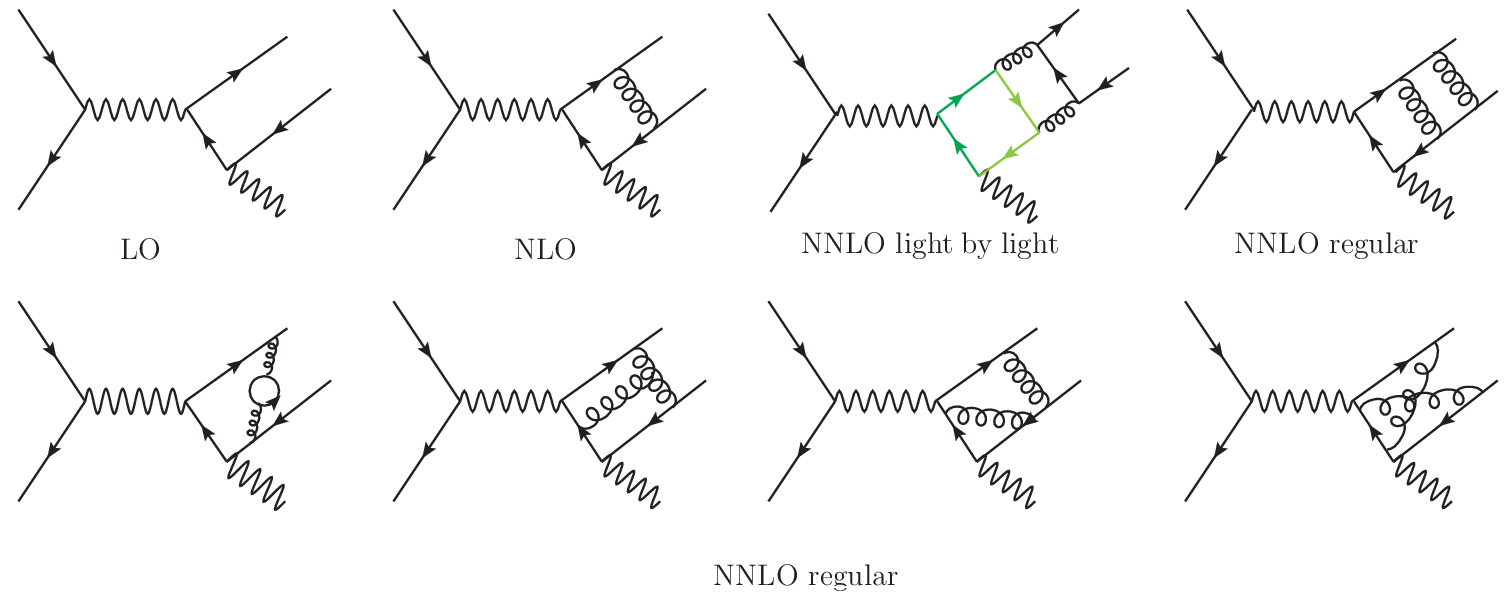}
\caption{Typical Feynman diagrams for the process $e^{+}+ e^{-}\to \gamma+\eta_c$ up to NNLO level, in which the outgoing $c$ and $\bar{c}$ quarks shall form the bound state $\eta_c$.}
\label{Fig0}
\end{figure}

As the first step, the package FeynArts~\cite{Hahn:2000kx} is used to generate the Feynman diagrams for the $\gamma+\eta_{c}$ production via the process $e^{+}+ e^{-}\to \gamma+\eta_c$ and the corresponding amplitudes up to NNLO level. There are totally $2$ LO Feynman diagrams, $8$ NLO Feynman diagrams and $120$ NNLO Feynman diagrams for the considered process, whose typical ones are presented in Fig.\ref{Fig0}. The evaluation involves the regular two-loop diagrams and the ``light-by-light" (lbl) scattering diagrams. The lbl graphs are both gauge-invariant and free of ultra-violet (UV) and infrared (IR) divergences. Second, we employ the packages FeynCalc/FormLink~\cite{Mertig:1990an, Feng:2012tk} to deal with the trace over the Dirac and color matrices. And then with the help of the packages Apart~\cite{Feng:2012iq} and FIRE~\cite{Smirnov:2014hma}, we conduct the partial fraction and the integration-by-parts reduction. Eventually, we end up with $6$ independent one-loop Master Integrals (MIs) and $174$ independent two-loop MIs. Those MIs can be evaluated through performing sector decomposition and subsequent numerical integrations with quadruple precision via the packages FIESTA/CubPack/HCubature~\cite{Smirnov:2013eza, CubPack, HCubature}. This explain our main procedures, and some more descriptions can be found in Ref.\cite{Feng:2019zmt}.

To implement the renormalization of the quark wave function and quark mass, we adopt the $\mathcal{O}(\alpha_{s}^{2})$-order on-shell renormalization constants $Z_{2}$ and $Z_{m}$, which expressions can be read from Refs.\cite{Broadhurst:1991fy, Melnikov:2000zc}. For self-consistency, the strong coupling constant under the $\overline{\rm MS}$ scheme is taken up to order-$\alpha_{s}^2$. After the renormalization procedure, the entire UV divergences are eliminated. To deal with infrared (IR) divergence, before conducting the loop integration, we expand the integrand of the quark amplitude in power series of $\mathbf{q^{2}}$, where $\mathbf{q}$ is the relative momentum between the two heavy constituent quarks. For this performance, the Coulomb singularities are automatically removed, thus we only need to consider the soft divergence. It is noted that the NNLO amplitude contains a single IR pole with the very coefficient as anticipated from the following equations (\ref{eq5}, \ref{eq6}, \ref{eq7}), as shown in the following illustrations, those IR terms can be explicitly factorized into the non-perturbative $\eta_c$ NRQCD matrix element according to $\overline{\rm MS}$ prescription, with $\log\mu_\Lambda$ manifested as the remnant ($\mu_\Lambda$ refers to the NRQCD factorization scale).

According to the NRQCD factorization formalism, the cross section for the process $e^{+} +e^{-} \to \gamma+\eta_c$ up to NNLO in $\alpha_s$ can be expressed as
\begin{eqnarray}\label{nrqcd-formalism}
\sigma=f^{(0)}\bigg[1+r_1a_s(\mu_r)+r_2(\mu_r)a_s(\mu_r)\bigg]\langle\mathcal{O}_1(^1S_0)\rangle,
\end{eqnarray}
where $a_s\equiv {\alpha_s}/{\pi}$, $f^{(0)}$ represents the LO short-distance coefficient (SDC), $r_1$ and $r_2$ correspond to the NLO and NNLO radiative corrections to the LO SDC, $\langle\mathcal{O}_1(^1S_0)\rangle$ denotes the NRQCD matrix element for the color-singlet and spin-singlet charmonium state. Since SDCs are insensitive to the nonperturbative hadronization effects, they can be deduced with the aid of the perturbative matching technique. That is, by replacing the physical $\eta_c$ meson with a fictitious onium composed of free $c\bar{c}$ pair, carrying the same quantum number as $\eta_c$. After this replacement, we have
\begin{eqnarray}\label{nrqcd-formalism-pert}
&& \sigma(c\bar{c}(^1S_0)) \nonumber\\
&=& f^{(0)}\bigg[1+r_1a_s(\mu_r)+r_2(\mu_r)a_s(\mu_r)\bigg]\langle\mathcal{O}_1(^1S_0)\rangle_{c\bar{c}(^1S_0)}.
\end{eqnarray}
Both the perturbative cross section $\sigma(c\bar{c}(^1S_0))$ and the NRQCD matrix element $\langle\mathcal{O}_1(^1S_0)\rangle_{c\bar{c}(^1S_0)}$ are calculable, and we can solve all the three SDCs $f^{(0)}$, $r_1$ and $r_2$. Concretely, we can employ the aforementioned techniques to evaluate perturbative cross section $\sigma(c\bar{c}(^1S_0))$ up to ${\mathcal O}(\alpha_s^2)$-order. Meanwhile, we are required to carry out the non-perturbative NRQCD matrix element at the two-loop level. Since the contributions from the potential region in the QCD calculation have been removed, to be consistent, we must remove this counterpart in our NRQCD evaluation. There still remain UV as well as IR divergences in NRQCD matrix element at the two loop level. The UV divergence can be renormalized in $\overline{\rm MS}$ scheme, thus, at the lowest order in velocity expansion, we have
\begin{eqnarray}\label{nrqcd-pert}
&& \bigg[\langle\mathcal{O}_1(^1S_0)\rangle_{c\bar{c}(^1S_0)}\bigg]_{\overline{\rm MS}} \nonumber\\
&&= 2 N_c\bigg[1-\alpha_s^2\frac{1}{\epsilon_{\rm IR}}\bigg(\frac{1}{2}C_F^2+\frac{1}{4}C_AC_F\bigg)\bigg],
\end{eqnarray}
which can be directly translated from Refs.\cite{Hoang:2006ty, Chung:2020zqc}. After hard work, we find that the remained divergence in $\sigma(c\bar{c}(^1S_0))$ can be exactly cancelled by the IR divergence in Eq.(\ref{nrqcd-pert}), which render the SDCs be free of any divergences, however develop a $\log \mu_\Lambda$ dependence.

In the following, we will directly present the SDCs and the cross section. The LO SDC reads
\begin{eqnarray}\label{LO-SDC}
f^{(0)}=\frac{32e_c^4\pi^2\alpha^3}{3m_cs^2}\bigg(1-\frac{4m_c^2}{s}\bigg).
\end{eqnarray}
Immediately, we obtain the cross section at $\mathcal{O}(\alpha_s^0)$
\begin{eqnarray}\label{LO-SDC}
\sigma^{(0)}=f^{(0)}\langle\mathcal{O}_1(^1S_0)\rangle.
\end{eqnarray}
For future convenience, we reexpress Eq.(\ref{nrqcd-formalism}) as
\begin{eqnarray}\label{cs}
\sigma=\sigma^{(0)}\bigg[1+r_1a_s(\mu_r)+r_2(\mu_r)a_s(\mu_r)\bigg],
\end{eqnarray}
where the tree-level cross section $\sigma^{(0)}$ is
\begin{equation}
\sigma^{(0)}=\frac{32e^{4}_{c}\pi^{2}\alpha^{3}}{3m_{c}s^{2}}\left(1-\frac{4m^{2}_{c}}{s}\right) \langle\mathcal{O}_{1}(^{1}S_{0}) \rangle,
\end{equation}
where $e_{c}=+2/3$ denotes the electric charge of charm quark, $m_{c}$ denotes the mass of charm quark, $\alpha$ is the fine structure constant, $\sqrt{s}$ is the center of mass energy, and $\langle\mathcal{O}_{1}(^{1}S_{0}) \rangle$ is the matrix element for $\eta_c$. The NLO coefficient $r_{1}$ takes the form~\cite{Sang:2009jc}
\begin{eqnarray}
r_{1} &=&-\frac{2[30q^{2}-(84+\pi^{2})q+2\pi^{2}+54]}{9(2-q)(1-q)} \nonumber\\
& & +\frac{8(2q-3)}{2(2-q)^{2}}\log{\left(\frac{2}{q}-2\right)} -\frac{4}{b}\log{\left(\frac{1-b}{1+b}\right)} \nonumber\\
&& + \frac{2}{3(q-1)}\left[\left(1+\frac{q}{2}\right)\log^{2}{(\frac{1-b}{1+b})} -\log^{2}{(\frac{2}{q}-1)}\right] \nonumber\\
&& +\frac{4}{3(1-q)}\textmd{Li}_{2}\left(\frac{q}{2-q}\right),
\end{eqnarray}
where $q={4m_{c}^{2}}/{s}$ and $b=\sqrt{1- {4m_{c}^{2}}/{s}}$.  The NNLO calculation technology has been described in detail in Ref.\cite{Feng:2015uha}. The explicit expressions for the NNLO coefficient $r_{2}$ is extremely lengthy.  For convenience, we define a $K$ factor,
\begin{equation}
K= \sigma/\sigma^{(0)}.
\end{equation}
Numerical results for the $K$ factor at the $e^+ e^-$ collision energy $\sqrt{s}=10.6$ GeV for three typical $m_c$ values are
\begin{widetext}
\begin{eqnarray}
K|_{m_{c}=1.4 \rm GeV}
&=&1-2.404 a_{s}(\mu_r) +[-80.11-37.29\log{\frac{\mu_{\Lambda}^{2}}{m_{c}^{2}}} +(-6.211+0.4007\;n_{l})\log{\frac{\mu_{r}^{2}}{m_{c}^{2}}}
\nonumber\\
&&-2.022\;n_{l}-3.852\;lbl ]a^{2}_{s}(\mu_r),  \label{eq5} \\
K|_{m_{c}=1.5 \rm GeV}
&=&1-2.569 a_{s}(\mu_r)+[-80.09-37.29\log{\frac{\mu_{\Lambda}^{2}}{m_{c}^{2}}} +(-6.635+0.4281\;n_{l})\log{\frac{\mu_{r}^{2}}{m_{c}^{2}}}
\nonumber\\
&&-1.959\;n_{l}-3.277\;lbl]a^{2}_{s}(\mu_r), ,  \label{eq6}\\
K|_{m_{c}=1.6 \rm GeV}
&=&1-2.722a_{s}(\mu_r)+[-80.19 -37.29\log{\frac{\mu_{\Lambda}^{2}}{m_{c}^{2}}} +(-7.032+0.4537\;n_{l})\log{\frac{\mu_{r}^{2}}{m_{c}^{2}}}
\nonumber\\
&& -1.893\;n_{l}-2.741\;lbl]a^{2}_{s}(\mu_r), \label{eq7}
\end{eqnarray}
\end{widetext}
where $\mu_{\Lambda}$ is the factorization scale, $n_{l}$ is the number of light flavors ($u$, $d$ and $s$), and ``$lbl$" denotes the contribution from the light-by-light Feynman diagrams. Though the $lbl$-terms are implicitly proportional to $n_l$, they are free of ultra-violet (UV) divergence and are irrelevant to the running of $\alpha_s$, so they should be kept as conformal terms when applying the PMC (Thus the $n_l$ in $lbl$-terms is fixed to be $3$). Using the above equations, the perturbative coefficients $r_{i}$ can be fixed. As a useful reference, we rewrite the $K$ factors for those typical $m_{c}$ values with explicit color structures in the following
\begin{widetext}
\begin{eqnarray}
K|_{m_{c}=1.4 \rm GeV}
&=&1-1.803\;C_{F}a_{s}(\mu_r)+[-4.935 (C_{A}C_{F} + 2 C_{F}^{2})\log{\frac{\mu_{\Lambda}^{2}}{m_{c}^{2}}} -0.4508 \beta_{0}C_{F}\log{\frac{\mu_{r}^{2}}{m_{c}^{2}}}
\nonumber\\
&&-6.599 C_{A}C_{F}-30.17 C_{F}^{2}-2.889C_{F}\;lbl -1.517 C_{F} n_l -0.0619C_{F}]a^{2}_{s}(\mu_r),
\\ K|_{m_{c}=1.5 \rm GeV}
&=&1-1.926 \;C_{F}a_{s}(\mu_r)+[-4.935 (C_{A}C_{F} + 2 C_{F}^{2})\log{\frac{\mu_{\Lambda}^{2}}{m_{c}^{2}}} -0.4816 \beta_{0}C_{F}\log{\frac{\mu_{r}^{2}}{m_{c}^{2}}}
\nonumber\\
&&-6.828C_{A}C_{F}-29.68C_{F}^{2}-2.457 C_{F}\;lbl-1.469 C_{F} n_l -0.0125C_{F}]a^{2}_{s}(\mu_r),
\\ K|_{m_{c}=1.6 \rm GeV}
&=&1-2.042\;C_{F}a_{s}(\mu_r)+[-4.935 (C_{A}C_{F} + 2 C_{F}^{2})\log{\frac{\mu_{\Lambda}^{2}}{m_{c}^{2}}} -0.5104 \beta_{0}C_{F}\log{\frac{\mu_{r}^{2}}{m_{c}^{2}}}
\nonumber\\
&&-7.049C_{A}C_{F}-29.28 C_{F}^{2}-2.056 C_{F}\;lbl -1.419 C_{F} n_l +0.0358 C_{F}]a^{2}_{s}(\mu_r),
\end{eqnarray}
\end{widetext}
where $C_F=4/3$ and $C_A=3$ are ${\rm SU}_{c}(3)$ color factors.

In order to apply the PMC, the perturbative coefficients $r_{i}$ need to be divided into conformal terms and non-conformal terms~\cite{Mojaza:2012mf, Brodsky:2013vpa}, and for the present NNLO analysis, the coefficients for the perturbative series (\ref{totcs}) can be written as
\begin{eqnarray}
r_{1} &=&r_{1,0}, \\
r_{2} &=&r_{2,0}+r_{2,1}\beta_{0},
\end{eqnarray}
where $\beta_0=11-2/3 n_f$ with the active flavor numbers $n_f=n_l+1$ ($n_f=4$ for $\eta_c$ production). The coefficients $r_{i,0}$ are the conformal coefficients and the $r_{i,j\neq 0}$ are non-conformal ones. By using the standard PMC scale-setting procedures, the NNLO total cross section (\ref{totcs}) can be written as a conformal series,
\begin{equation}
\sigma=\sigma^{(0)}[1+ r_{1,0} a_s(Q_\star) +r_{2,0}a^2_s(Q_\star)],
\label{ten}
\end{equation}
where $Q_\star$ stands for the PMC scale which is determined by requiring all non-conformal terms to vanish. The PMC scale can be fixed up to leading-logarithm (LL) accuracy by using the known NNLO pQCD series;  i.e.,
\begin{equation}
\ln{\frac{Q_{\star}^{2}}{m_{c}^{2}}}=-\frac{\hat{r}_{2,1}}{r_{1,0}} + {\cal O}(a_s),
\label{elven}
\end{equation}
where $\hat{r}_{2,1}=r_{2,1}|_{\mu_{r}=m_{c}}$.

The PMC scale $Q_{\star}$ can be regarded as the effective momentum flow of the process, since it is determined by using the RGE and the effective value of $\alpha_s(Q_\star)$ has been fixed. Eq.(\ref{elven}) shows that the scale $Q_{\star}$ is independent of the choice of $\mu_r$. Since the conformal coefficients $r_{i,0}$ are scale-independent, the PMC prediction is exactly free of $\mu_r$-dependence. Thus the conventional renormalization scale ambiguity is solved.

The unknown higher-order terms in the perturbative series of $\ln{{Q_{\star}^{2}}/{m_{c}^{2}}}$ can have some residual scale dependence~\cite{Zheng:2013uja}; however, this dependence is distinct from the conventional renormalization scale ambiguities, and it has both  $\alpha_s$-power suppression and exponential suppression. In practice, most applications of the PMC published in the literature show that such residual scale dependence is rather small~\footnote{In some cases where both the resulting conformal series and the perturbative series of the PMC scale do not converge well, the residual scale dependence may be large. This dependence is expected to be suppressed when one includes higher loop terms. An example can be found for Higgs-boson decay $H \to gg$~\cite{Zeng:2018jzf}.}~\cite{Wu:2013ei, Wu:2019mky}.

\section{Numerical Results}

To do the numerical calculation, we take $\alpha$=${1}/{130.7}$, $\alpha_{s}(M_{Z})$=0.1181~\cite{Tanabashi:2018oca}, $\sqrt{s}=10.6$ GeV, and the $\eta_c$ matrix element $\langle\mathcal{O}_{1}(^{1}S_{0}) \rangle|_{\mu_\Lambda=1{\rm GeV}}$=0.437 $\rm GeV^{3}$~\cite{Bodwin:2007fz, Chung:2010vz}. This value is fixed via a matching at the NLO level without factorization scale dependence. At present, the NNLO evolution equation of the matrix element is missing and we adopt the one-loop evolution equation~\cite{Bodwin:1994jh} to fix the matrix element at different scales, there is thus factorization scale dependence for the NNLO total cross section (\ref{totcs}). For definiteness, we have explicitly set the factorization scale of the matrix element as the usually adopted one, i.e. $\mu_\Lambda=1{\rm GeV}$. If setting its scale as another usual choice, e.g. $\mu_\Lambda=m_c$, we find that the total cross-section shall only be slightly changed, e.g. less than $5\%$ of the total cross-section. The required two-loop $\alpha_s$ running is calculated by using the RunDec program~\cite{Chetyrkin:2000yt}.

Assuming $m_{c}=1.5$ GeV and the factorization scale $\mu_\Lambda=1$ GeV,  the total cross sections under conventional scale-setting approach (Conv.) for three typical choices of renormalization scale are
\begin{eqnarray}
\sigma^{\rm Conv}_{\rm NNLO}\left|_{\mu_r=\sqrt{s}/2}\right. &=& 0.51\sigma^{(0)}, \\
\sigma^{\rm Conv}_{\rm NNLO}\left|_{\mu_r=\sqrt{s}} \right.    &=& 0.61\sigma^{(0)}, \\
\sigma^{\rm Conv}_{\rm NNLO}\left|_{\mu_r=2\sqrt{s}} \right.    &=& 0.68\sigma^{(0)},
\end{eqnarray}
which shows that the conventional scale uncertainty is about $\left(^{+11.5\%}_{-16.3\%}\right)$ for $\mu_r \in[\sqrt{s}/2, 2\sqrt{s}]$.
On the other hand, since the PMC scale $Q^*$ is fixed to be $14.78$ GeV, the total cross section using PMC scale-setting approach is independent of the choice of $\mu_r$; i.e.
\begin{equation}
\sigma^{\rm PMC}_{\rm NNLO} \equiv 0.64\sigma^{(0)}.
\end{equation}

\begin{table}[htb]
\centering
\begin{tabular}{ccccc}
\hline
      $K$ factor & $\alpha_s^0$-terms & $\alpha_s^1$-terms & $\alpha_s^2$-terms & Total  \\
\hline
      Conv. & 1 & $-0.14^{+0.02}_{-0.03}$ & $-0.25^{+0.05}_{-0.07}$ & $0.61^{+0.07}_{-0.10}$   \\
\hline
      PMC & 1 & -0.13 & -0.23 & 0.64  \\
\hline
\end{tabular}
\caption{The magnitudes of various perturbative terms to the NNLO $K$ factor using conventional and PMC scale-setting approaches, respectively. $\mu_\Lambda=1$ GeV. The central values are for $\mu_r=\sqrt{s}$ and the errors are for $\mu_r \in[\sqrt{s}/2, 2\sqrt{s}]$ for conventional scale-setting. The PMC predictions are independent of the choice of $\mu_r$. }
\label{Tableone}
\end{table}

Table \ref{Tableone} shows how the magnitude of each loop term to the NNLO $K$ factor changes under different choices of $\mu_r$. After applying the PMC, the scale dependence of each loop term is eliminated; however, the pQCD convergence is still poor. This is due to the fact that the conformal coefficient is large and dominates over the total NNLO coefficient $r_2$, e.g. $r_{2}=-79.9\pm 7.4$ where $r_{2,0}\equiv -83.5$ for $\mu_r\in[\sqrt{s}/2, 2\sqrt{s}]$. We have observed numerically that if one sets $\mu_r=\sqrt{2s}$, the conventional prediction gives  total cross section and perturbative behavior which are close in comparison to the PMC prediction; thus $\sqrt{2s}$ can be treated as the optimal renormalization scale of conventional prediction.

\begin{figure}[htb]
\centering
\includegraphics[width=0.5\textwidth]{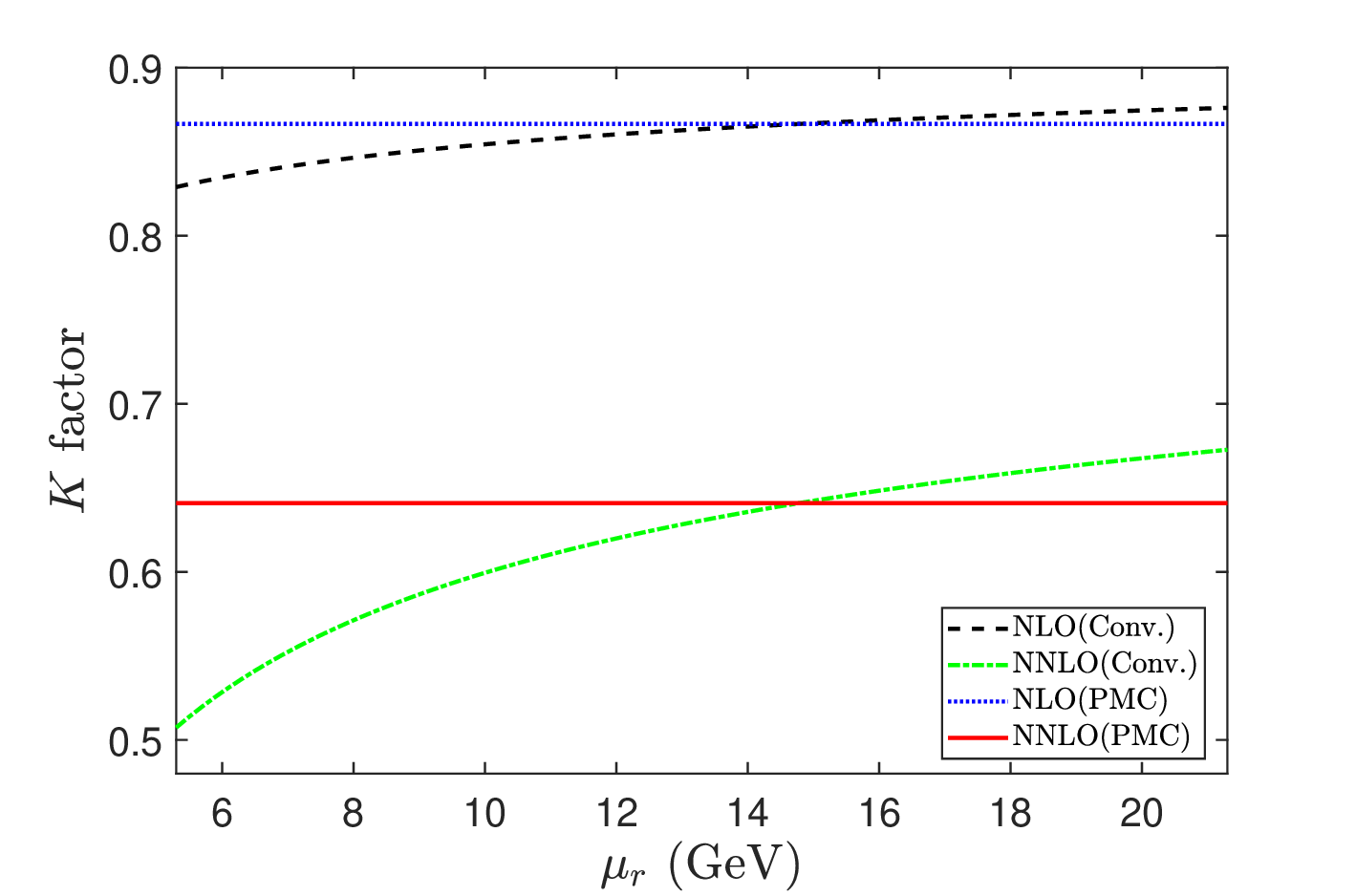}
\caption{The $K$ factors versus the renormalization scale $\mu_{r}$ up to NLO and NNLO levels under conventional and PMC scale-setting approaches, respectively. $\mu_\Lambda=1$ GeV. }
\label{Fig1}
\end{figure}

Fig.(\ref{Fig1}) shows how the net scale uncertainty varies when more loop terms have been included. Contrary to usual expectations, the net NNLO scale uncertainty of the conventional series is larger than that of the NLO series, since cancellations of terms at different orders are absent. In contrast, after applying the PMC, both the NLO and NNLO $K$ factors are free of the renormalization scale ambiguities. In this sense, the scale-invariant PMC conformal series is extremely important for precise pQCD predictions.

After eliminating the renormalization scale uncertainty, there are other error sources such as the factorization scale, the charm quark mass $m_c$, the value of $\alpha_s(M_Z)$, and the $\eta_c$ matrix element $\langle\mathcal{O}_{1}(^{1}S_{0})\rangle$.  The matrix element is an overall parameter, whose error can be determined separately. Thus, in the following, we shall only discuss uncertainties which come from $\mu_\Lambda$, $m_c$ and $\Delta\alpha_s(M_Z)$. When discussing the uncertainty of the PMC prediction from one parameter, the other parameters are set as their central values.

\begin{figure}[htb]
\centering
\includegraphics[width=0.5\textwidth]{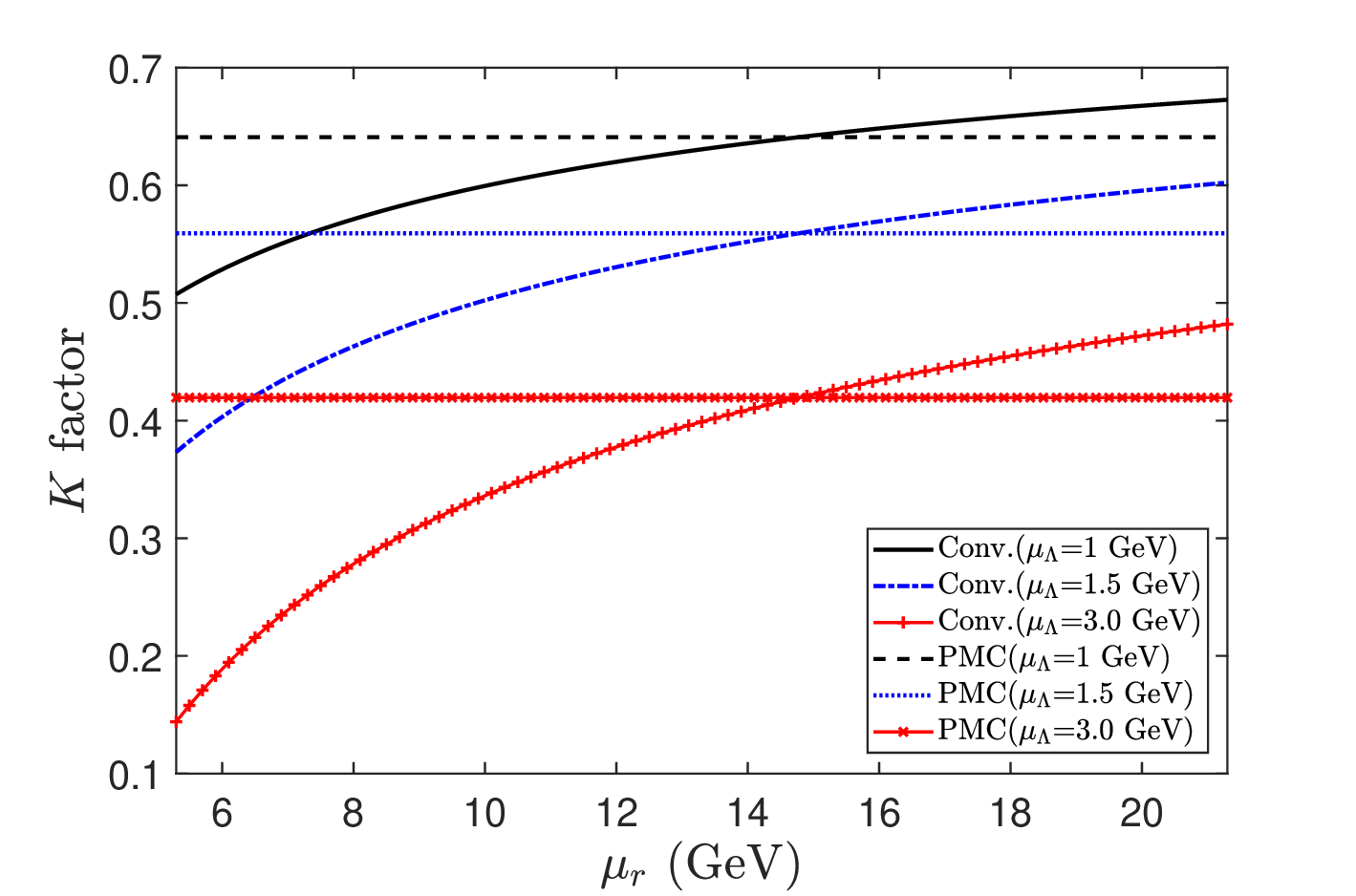}
\caption{The NNLO $K$ factors versus $\mu_{r}$ under conventional and PMC scale-settings. Three factorization scales 1 GeV, $m_{c}$, $2m_{c}$ are adopted. $m_{c}=1.5$ GeV. }
\label{Fig2}
\end{figure}

First, to discuss the factorization scale uncertainty, we adopt $\mu_{\Lambda}$=1 GeV, $m_{c}$ and $2m_{c}$ with $m_c=1.5$ GeV, respectively. The $K$ factors are
\begin{equation}
K^{\rm PMC}={0.64,\;\; 0.56,\;\; 0.42},
\end{equation}
accordingly. Fig.(\ref{Fig2}) shows that the factorization scale dependence under conventional and PMC scale-setting approaches, respectively. It shows that the $K$ factor decreases with increasing factorization scale.

\begin{figure}[htb]
\centering
\includegraphics[width=0.5\textwidth]{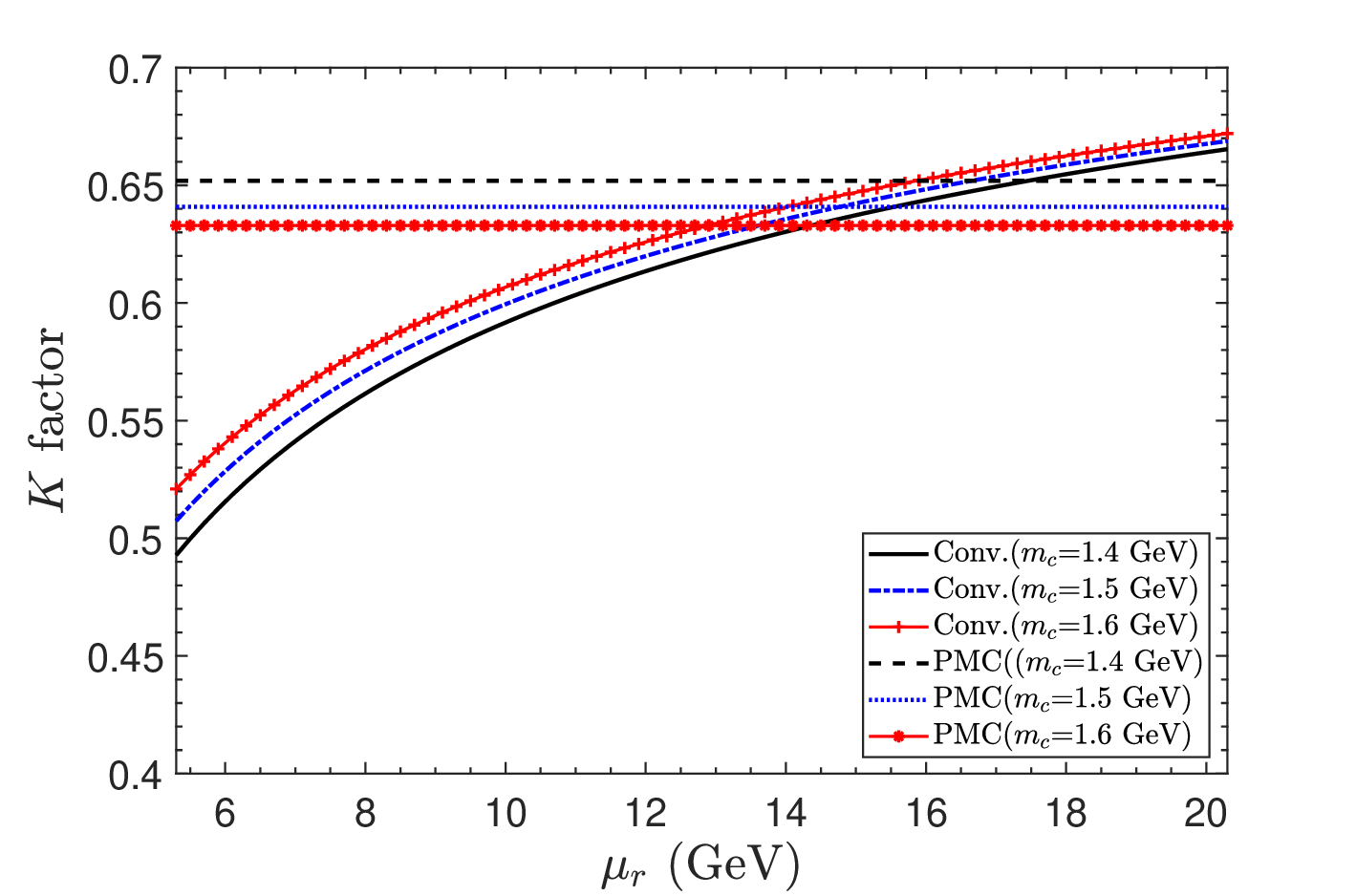}
\caption{The NNLO $K$ factors versus $\mu_{r}$ under conventional and PMC scale-settings, where $m_{c}=1.4$, $1.5$ and $1.6$ GeV, respectively. $\mu_\Lambda=1$ GeV.}
\label{Fig3}
\end{figure}

Second, to discuss the $m_c$ uncertainty, we adopt $m_c=1.4$, $1.5$, and $1.6$ GeV, respectively. By setting $\mu_\Lambda=1$ GeV, we obtain
\begin{equation}
K^{\rm PMC}={0.65,\;\; 0.64,\;\; 0.63},
\end{equation}
accordingly. Fig.(\ref{Fig3}) shows that the $m_c$ dependence under conventional and PMC scale-setting approaches, respectively. The $K$ factor decreases with the increment of $m_c$.

\begin{figure}[htb]
\centering
\includegraphics[width=0.5\textwidth]{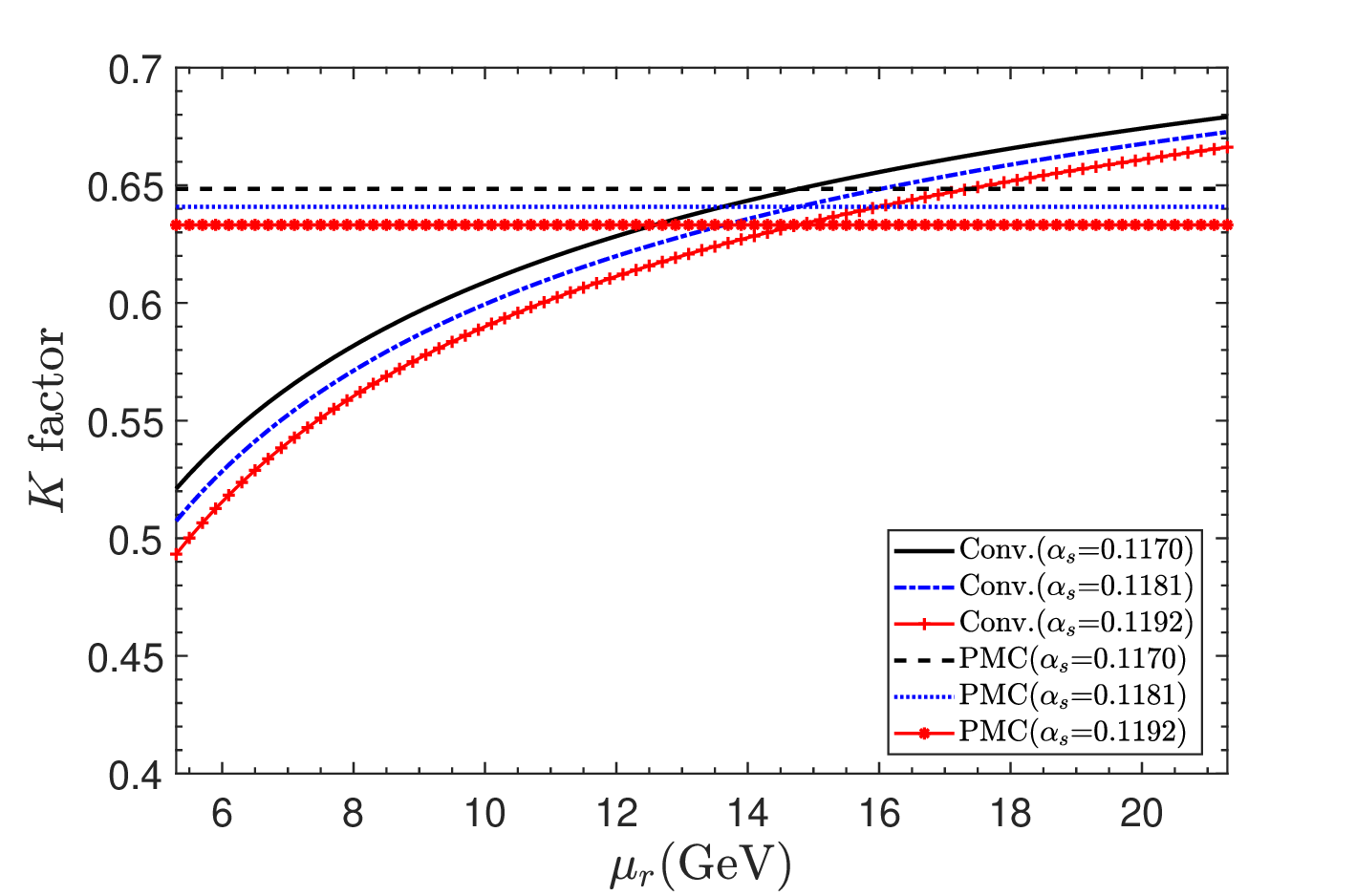}
\caption{The NNLO $K$ factors versus $\mu_{r}$ under conventional and PMC scale-settings, where $\Delta\alpha_{s}(M_{Z})=\pm0.0011$. $m_c=1.5$ GeV and $\mu_\Lambda=1$ GeV. }
\label{Fig4}
\end{figure}

Third, to discuss the $\Delta\alpha_{s}(M_{Z})$ uncertainty, we adopt $\Delta\alpha_{s}(M_{Z})=0.1181\pm0.0011$~\cite{Tanabashi:2018oca}, and we obtain
\begin{equation}
K^{\rm PMC}={0.65,\;\; 0.64,\;\; 0.63},
\end{equation}
accordingly.  Fig.(\ref{Fig4}) shows that the $\Delta\alpha_{s}(M_{Z})$ dependence under conventional and PMC scale-setting approaches, respectively.

We are now ready to discuss the properties of the total cross section. The PMC NNLO total cross section is
\begin{eqnarray}\label{PMCnnlo}
\sigma^{\rm PMC}_{\rm NNLO} &=& 41.18^{+8.17+4.76+0.72}_{-11.83-3.92-0.73} \;\; {\rm fb},
\end{eqnarray}
where the first error is for $\mu_{\Lambda}\in[1, 3]$ GeV, the second error is for $m_c\in[1.4, 1.6]$ GeV, and the third error is for $\Delta\alpha_s(M_{Z})=\pm0.0011$. The central value of the total cross-section is for $\mu_\Lambda=1.5$ GeV, $\alpha_{s}(M_{Z})=0.1181$, and $\langle \mathcal{O}_{1}(^{1}S_{0})\rangle|_{\mu_\Lambda=1.5{\rm GeV}}=0.418 \;{\rm GeV}^3$, which is obtained by evolving the matrix element $\langle \mathcal{O}_{1}(^{1}S_{0})\rangle|_{\mu_\Lambda=1.0{\rm GeV}}=0.437\; {\rm GeV}^3$ from 1 GeV to 1.5 GeV with the help of the one-loop evolution formulae given in Ref.\cite{Bodwin:1994jh}.

Recently, the BELLE Collaboration published their measured total cross-section for $e^{+}+e^{-}\to\eta_c+\gamma$ at $\sqrt{s}$=10.58 GeV. There measured Born cross section $\sigma_{\rm B}$=$11.3^{+7.0+1.5}_{-6.6-1.5}$ fb~\cite{Jia:2018xsy}, and by employing the following formula~\cite{Dong:2017tpt}
\begin{equation}
\sigma_{\rm B}(s)=(1+\delta(s))\frac{\sigma_{\rm obs}(s)}{1/|1-\prod(s)|^2},
\end{equation}
we inversely obtain $\sigma_{\rm obs}$=$16.58^{+10.51}_{-9.93}$ fb. This value is smaller than the theoretical prediction (\ref{PMCnnlo}). The central theoretical value shows a  2.3$\sigma$ deviation from the data, which increases to 3.1$\sigma$ when taking the usual choice of $\mu_\Lambda=1$ GeV.  If one  takes the uncertainty of the $\eta_c$ matrix element into consideration, e.g. $\Delta\langle \mathcal{O}_{1}(^{1}S_{0}) \rangle|_{\mu_\Lambda=1.5{\rm GeV}}=\left(^{+0.111}_{-0.105}\right) {\rm GeV}^3$~\cite{Bodwin:2007fz, Chung:2010vz}, we then have an  additional uncertainty, $\left(^{+10.94}_{-10.34}\right)$ fb, to the total cross section. This gives a slight overlap of  the theoretical prediction with the measurement.

As a final remark, as shown by Table~\ref{Tableone}, the poor pQCD convergence is due to the intrinsic nature of the present process, which cannot be improved even after applying the PMC. Thus it is necessary to know the contributions from unknown terms before we draw any definite conclusions. The conventional error estimate obtained by varying the guessed scale over a certain range cannot give a reliable prediction of the unknown terms, since it only partly estimates the non-conformal contribution but not the conformal one.

If one has a renormalization-scale independent conformal series, one can often obtain a reliable prediction of unknown higher-order contributions~\cite{Du:2018dma} with the help of the $\rm Pad\acute{e}$ resummation~\cite{Basdevant:1972fe, Samuel:1992qg, Samuel:1995jc}. The diagonal [1/1]-type $\rm Pad\acute{e}$ series is generally preferable for estimating the unknown contributions from a poor pQCD convergent series~\cite{Gardi:1996iq, Cvetic:1997qm, Yu:2019mce}; and for the present process, the poor convergence of PMC series is due to large conformal coefficients. Detailed procedures for obtaining a combined $\rm Pad\acute{e}$  +PMC prediction can be found in Ref.\cite{Yu:2019mce}. By using the diagonal [1/1]-type PAA approximant, the predicted coefficient of the $\rm N^{3}LO$ term is -2713.77, which results in an extra $38\%$ suppression from the LO cross-section, leading to a NNNLO prediction in better agreement with the data;  i.e.,
\begin{eqnarray}
\sigma_{[1/1], {\rm NNNLO}}^{\rm PMC} =21.36\;\; {\rm fb},
\end{eqnarray}
where $\mu_\Lambda=1$ GeV and the other parameters are set to their central values. This indicates the importance of a strict NNNLO calculation, even though it would be much more difficult than the present NNLO calculation. \\

\section{Summary}

In this paper, we have presented a detailed study of the cross section for $\gamma+\eta_c$ production in electron-position collisions up to NNLO level. If one used the conventional procedure, the renormalization scale uncertainty for the NNLO total cross-section is estimated as $28\%$ by varying the scale within the the range of $\mu_r \in[\sqrt{s}/2, 2\sqrt{s}]$. However, after applying PMC scale setting, the conventional scale uncertainty is eliminated. Our NNLO prediction is,
\begin{equation}
\sigma^{\rm PMC}_{\rm NNLO}=41.18^{+9.48}_{-12.48}\;\; {\rm fb},
\end{equation}
where the errors are squared average of the errors caused by varying $\mu_{\Lambda}\in[1, 3]$ GeV, $m_c\in[1.4, 1.6]$ GeV, and $\Delta\alpha_s(m_Z)=\pm0.0011$. Among the uncertainties from the other input parameters, the factorization scale error is the largest. The central value of the NNLO cross section deviates substantially from the measured data. The poor pQCD convergence of the series indicates the importance of uncalculated NNNLO terms for this process.  An initial estimate of the NNNLO terms with the help of the PMC and $\rm Pad\acute{e}$ resummation has been given in Section III;  the magnitude of the NNNLO contribution is sufficiently large to explain the data. Even though we need more accurate data to confirm the theoretical results,  this application of the PMC shows the importance of a correct renormalization scale setting for a reliable pQCD prediction. \\

\noindent{\bf Acknowledgement}: This work is partly supported by the Chongqing Graduate Research and Innovation Foundation under Grant No.ydstd1912, the National Natural Science Foundation of China under Grant No.11625520, No.11975187 and No.11947406, and the Fundamental Research Funds for the Central Universities under Grant No.2020CQJQY-Z003. The work of SJB is supported by the Department of Energy, Contract No. DE-AC02- 76SF00515. OSTI ID = 1643687.

\end{document}